**TITLE**: Single-Step Synthesis of Shape-Controlled Polymeric Particles using Initiated Chemical Vapor Deposition in Liquid Crystals

**Short Title:** Single-Step Synthesis of Shaped Polymeric Particles

**Authors:** Apoorva Jain, Soumyamouli Pal, Nicholas L. Abbott*, Rong Yang*

*Robert Frederick Smith School of Chemical and Biomolecular Engineering, Cornell University*

**Abstract:**

The ability to synthesize shape-controlled polymer particles will benefit a wide range of applications including targeted drug delivery and metamaterials with reconfigurable structures, but existing synthesis approaches are commonly multistep and limited to a narrow size/shape range. Using a novel single-step synthesis technique, a variety of shapes including nanospheres, hemispherical micro-domes, orientation-controlled microgels, microspheres, spheroids, and micro-discs were obtained. The shape-controlled particles were synthesized by polymerizing divinylbenzene (DVB) via initiated chemical vapor deposition (iCVD) in nematic liquid crystals (LC). iCVD continuously and precisely delivered vapor-phase reactants, thus avoiding disruption of the LC structure, a critical limitation in past LC-templated polymerization. That shape controllability was further enabled by leveraging LC as a real-time display of the polymerization conditions and progression, using a custom in-situ long-focal range microscope. Detailed image analysis unraveled key mechanisms in polymer synthesis in LC.

Poor solubilization by nematic LC led to the formation of pDVB nanospheres, distinct from microspheres obtained in isotropic solvents. The nanospheres precipitated to the LC-solid interface and further aggregated into microgel clusters with controlled orientation that was guided by the LC molecular alignment. On further polymerization, microgel clusters phase separated to form microspheres, spheroids, and unique disc-shaped particles.

**Teaser:** Shaped polymeric particles were synthesized in a single step via iCVD by leveraging Liquid Crystals as a template and real-time display.

**MAIN TEXT:**

**INTRODUCTION**

Polymer microparticles have achieved enormous impact on the global economy, with applications ranging from drug delivery in oncology[1] to enabling precision separations in chromatography.[2–4] Recent fundamental advances have further revealed how particle shape determines their packing and assembly, pointing to metamaterials with reconfigurable structures and programmable dynamic behaviors,[5,6] and how shape dictates the fate of microparticles in a human body.[7–12] The need for shape-controlled polymer particles has spurred rapid development of new synthesis approaches, ranging from microfluidic devices,[13,14] to seeded emulsion polymerization,[15,16] to particle replication in non-wetting templates (PRINT),[17] to mechanical stretching.[18–24] While these techniques have led to exciting new shapes, their shape formation fundamentally relies on physical manipulations, which are laborious and unlikely scalable. There have been few fresh conceptual advances beyond emulsion polymerization and physical manipulation in achieving shape control in recent years.

Take poly(divinylbenzene) (pDVB) microspheres as an example, the synthesis of which has been extensively studied in precipitation polymerization,[25] owing to its commercial use as chromatographic packing beads. As unveiled by Stover et al,[25–31] pDVB microspheres form under marginal solvency, via a 5-step growth mechanism. Starting with a solution of DVB monomers in a marginal solvent (e.g., acetonitrile), oligomeric polymer chains form upon initiation of polymerization, which then aggregate into colloidally-stable microgel nuclei as polymerization proceeds. The microgel nuclei eventually phase separate as solid microparticles and grow further by capturing oligomeric chains through pendant vinyl bonds at their surfaces. The final size is controlled by the solution composition (e.g., concentrations of monomer and initiator) and typically falls in the range of 3 – 5 µm.[25] The as-formed microparticles are uniformly spherical due to the dominance of surface tension during the phase separation in solution.

Here, we report a new framework for understanding the synthesis of shape-controlled polymer particles in liquid crystals (LC), one that is based on the discovery that DVB polymerization occurs at two distinct locations in an anisotropic solvent. Our work builds upon the exciting finding that LCs, with both fluidity and long-range order, can simultaneously serve as a solvent and a template during polymerization. The templating effect gives rise to anisotropic polymeric networks [32–35] and oriented nanofibers.[36–38] While the long-range order directs shape formation, the elastic properties of LC are proposed to influence the dimensions of the polymeric structures. Nevertheless, the molecular mechanisms that led to the striking morphologies obtained in LC remain elusive, limiting the capability to rationally design polymeric structures using structured liquids as templates.

This report bridges that knowledge gap by establishing effective real-time monitoring of the microscopic progression of polymerization, leveraging the optical output inherent to LC as indicators of the reacting microenvironment, and the continuous delivery of reactants enabled by initiated Chemical Vapor Deposition (iCVD) (**Figure 1**A-B). Drawing inspiration from the reaction mechanism of precipitation polymerization under marginal solvency, we reveal that polymer morphologies emerge from a two-stage growth pathway, each resembling the 5-step growth mechanism identified by Stover but at different locations in the LC (Figure 1D). In bulk LC, the 5-step polymerization yields phase-separated nanospheres in the size range of 160 ± 80 nm. Upon precipitation at the LC-substrate interface, the nanospheres aggregate while capturing monomer/oligomers via heterogenous polymerization to form microgel clusters. The aspect ratio and orientation of the clusters are directed by the anisotropic elastic stress of the LC templates near that interface. Upon further aggregation/polymerization, micron-sized spheres, spheroids, and discs emerge from the clusters via phase separation. These new insights into the mechanism of particle shape formation in LCs enable precise control over the polymer morphology, expanding it from polymeric networks [32–35] and nanofibers[36–38] to discrete and shape-controlled nano/microparticles. The reported approach could lead to drug delivery systems with programmable pharmacokinetics[7,9,12] and targeting effects or self-propelled microrockets with controlled group behaviors.[5] The fundamental insights obtained in this work have the potential to inform the development of novel synthesis mechanisms, applicable to precipitation polymerization, emulsion polymerization, and other solution polymerization approaches.

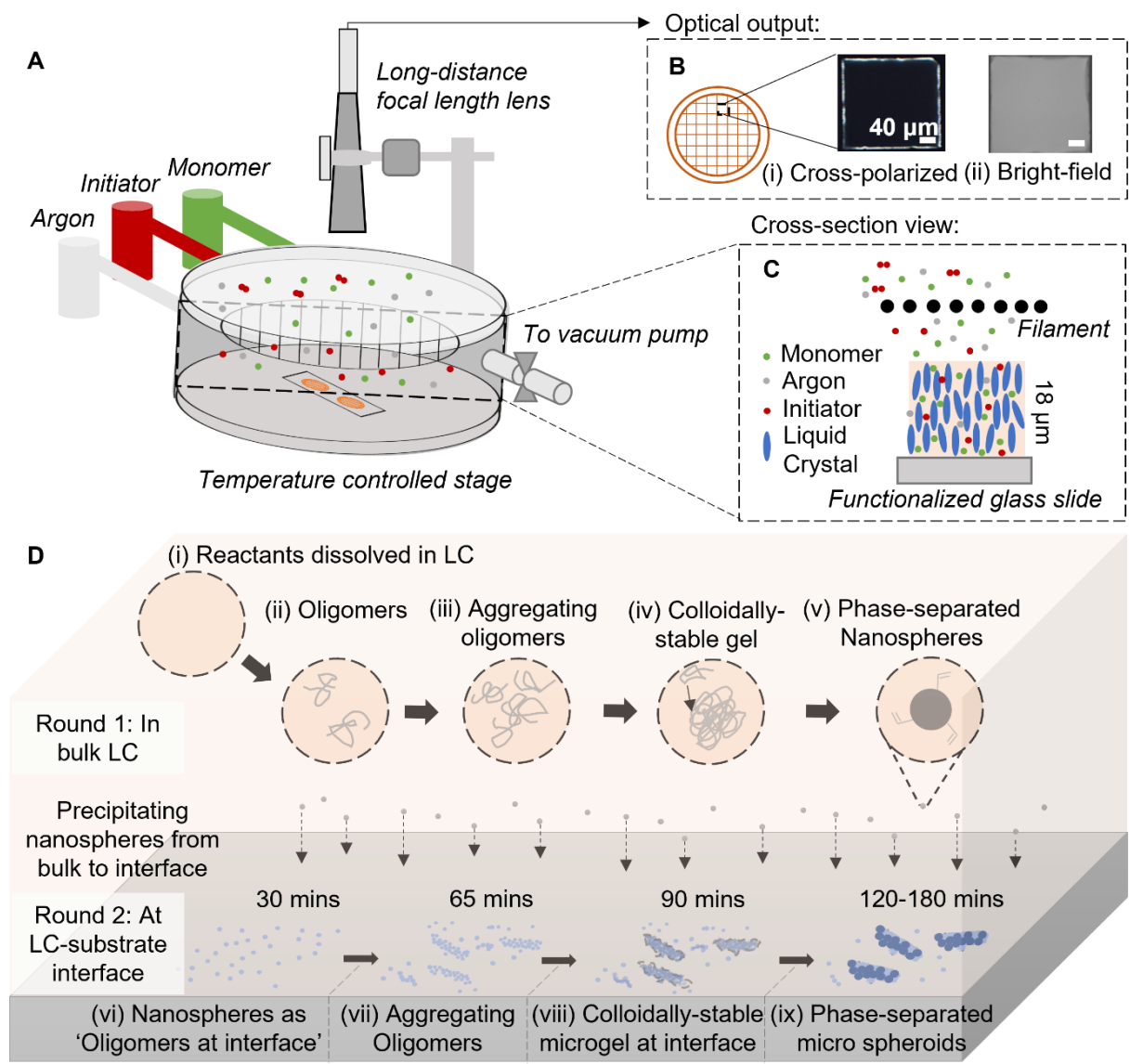

**Figure 1. (A) Schema of the iCVD synthesis apparatus, (B) two examples of the optical output monitored in real time using the custom in-situ long-focal range microscope and, (C) cross-section schema of the iCVD reactor chamber,** where initiator breaks down to form free radicals. The free radicals enter the LC and start polymerization upon reacting with dissolved monomers, giving rise to the polymeric particles with varying shapes. **(D) Mechanistic insights into the iCVD in LC process.** Bulk nematic LC acts as a marginal solvent resulting in continuous precipitation of pDVB as nanospheres. The nanospheres form via the 5-step precipitation polymerization mechanism: (i) dissolution of monomer in the LC, (ii) formation of oligomers, (iii) particle nucleation via coagulation of oligomers, (iv) growth

into colloidally stable seed and, (v) phase-separation into nanosphere. The LC-substrate interface facilitates a second round of the 5-step growth, where the precipitated nanospheres at the interface (analogous to the oligomeric building blocks in bulk) form particle nuclei, LC-directed microgel clusters, and then phase-separated micron-sized shaped particles.

**RESULTS**

**iCVD polymerization in LC and its real time monitoring**

A fundamental limitation of LC-templated polymerization has been the disruption of the LC template structure by loading monomers into the LC at a concentration sufficient to sustain the polymerization.[37,39] To minimize perturbation of the LC phase by monomer dissolution, we chose iCVD based on its continuous and precise vapor-phase delivery of monomers. Furthermore, iCVD is compatible with a large library of functional monomers, enabling access to a wide palette of polymers with diverse chemical functionalities. While iCVD is presently known for its conformal coating ability over solid substrates, Gupta et al[40] have used iCVD in isotropic solvents such as ionic liquids and silicone oil to form 3D polymeric structures. For insoluble monomers like 1H,1H,2H,2H-perfluorodecyl acrylate (PFDA), polymerization occurred predominantly at the vapor-liquid interface; while soluble monomers such as 2-hydroxyethyl methacrylate (HEMA) infiltrate the bulk liquid, giving rise to a liquid-polymer gel.[41]

During iCVD in LC (Figure 1A), the monomer DVB and initiator *tert*-butyl peroxide (TBPO) are vaporized and delivered at controlled rates into the reactor under medium-to-high vacuum (150 mTorr, see Supporting Information Table S1 for detailed deposition conditions). TBPO is thermally decomposed to form *tert*-butoxide (TBO) radicals, with the energy supplied by a filament array that is resistively heated to ~270ºC. The array is suspended over a cooled stage (kept at 20ºC), on which the nematic LC, E7, is placed. The DVB vapor and TBO radicals partition into the LC, which subsequently react to initiate polymerization via the chain-growth mechanism. To ensure that the dissolution of DVB and TBO, and the polymerization of DVB did not disrupt the alignment of the LC medium, we leveraged our custom long focal range microscope to observe the real-time progression of these processes (**Figure 2**).

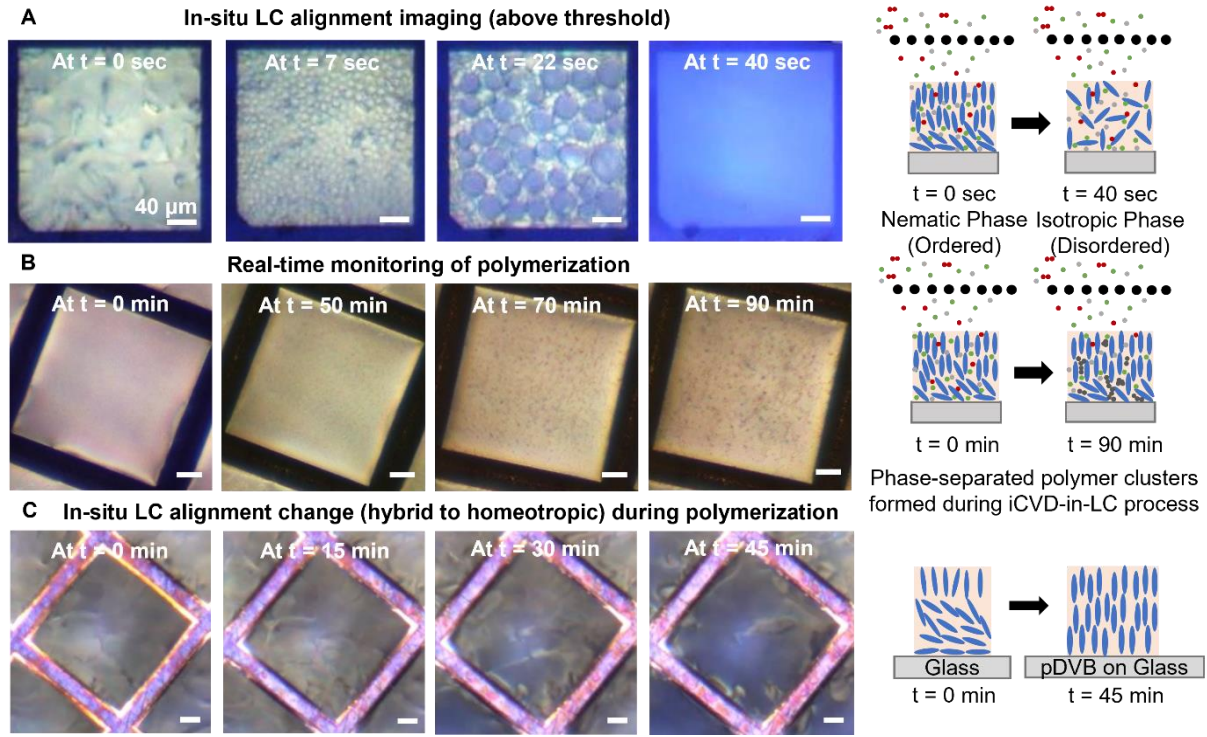

**Figure 2. In-situ characterization during iCVD in LC enabled by our custom long focal range microscope.** (**A**) Determination of monomer threshold concentration by observing LC phase transition from nematic to isotropic phase, (**B**) In-situ observation of emergence of phase separated polymer structures during iCVD polymerization in LC and, (**C**) Real-time monitoring of LC alignment captures lowering of LC anchoring energy during iCVD polymerization.

We first identified the upper bound of DVB concentration in the LC template, above which the nematic LC transitioned to an isotropic phase with no long-range orientational order, as demonstrated from a series of snapshots taken in-situ (Figure 2A and Supporting Information, iCVD polymerization was performed in the nematic LC by co-flowing DVB and TBPO (see Supporting Information Table S1 for detailed flowrates, temperatures, and pressure during iCVD). 50 minutes after the start of the reaction (by turning the filament array to decompose TBPO into radicals), we observed formation of phase separated polymeric clusters (dark dots in Figure 2B and Supporting Information, Movie S2) in LC (the beige background). We later confirmed that these dark dots correspond to polymeric clusters of different shapes using ex-situ microscopy images and SEM, as described in section 2.2.

Movie S1). The orientational order of LC was determined using cross polarizers in the in-situ microscope. The bright appearance at $t = 0$ indicated the hybrid orientation of LC that is typically observed over a glass substrate (Figure 2A schematic on the left),[42] whereas the blue color ($t = 40$ seconds) indicates an isotropic liquid (Figure 2A schematic on the right; the blue color was due to light scattering from the bottom of the reactor when viewed using crossed polars). As such, we identified the threshold DVB concentration, expressed as the DVB partial pressure in the vapor phase directly above the LC, i.e., $P_{DVB}$, to be 30.5 mTorr (see Supporting Information Table S2 for details). The nematic to isotropic transition was signified by formation of isotropic droplets that we observed to commonly accompany the phase transition. The $P_{DVB}$ of 26.9 mTorr was used for all subsequent experiments. This partial pressure was used to ensure that LC stays in the nematic phase to provide the templating effect, while achieving a high rate of polymerization. For the initiating species, i.e., TBPO and TBO radicals, no such transition was observed for $P_{TBPO}$ values up to 90 mTorr. $P_{TBPO}$ of 3.8 mTorr was used throughout this study.

The real-time monitoring also revealed a surprising effect of the polymerization, i.e., a transition of the LC alignment (on glass) from hybrid (bright appearance) to homeotropic (dark appearance) over the course of 45 minutes of DVB polymerization (Figure 2C and Supporting Information, Movie S3). We confirmed the transition by characterizing LC films obtained before (hybrid) and after (homeotropic) 65 minutes of iCVD using ex-situ orthoscopic and conoscopic polarized light microscopy (see Supporting Information, Figure S1). To unravel the origin of the transition, we prepared an optical cell with two glass slides coated using the LC-templated iCVD process, capping the top and bottom of a E7 film (see Supporting Information, Figure S1C). We argue that if the orientational transition had occurred via reorientation of the easy axis (preferred orientation) of the LC at the confining surfaces (e.g., due to polymeric structure at a surface) then the LC would maintain its perpendicular alignment in the optical cell (with dark appearance under cross-polarized light). However, the optical cell showed a bright texture (Figure S1F) corresponding to a planar alignment (Figure S1I). Therefore, we attributed the transition in Figure 2C to a low LC surface anchoring energy generated by the iCVD polymerization at the LC-substrate interface, such that the anchoring energy at the air-LC interface dictates the homeotropic alignment of the entire LC film, as shown in Figure 2C. This guided our investigation into the LC-substrate interface and our discovery of the role it plays in particle shape evolution.

**Discovery of shape evolution during iCVD polymerization in LC and elucidation of mechanisms**

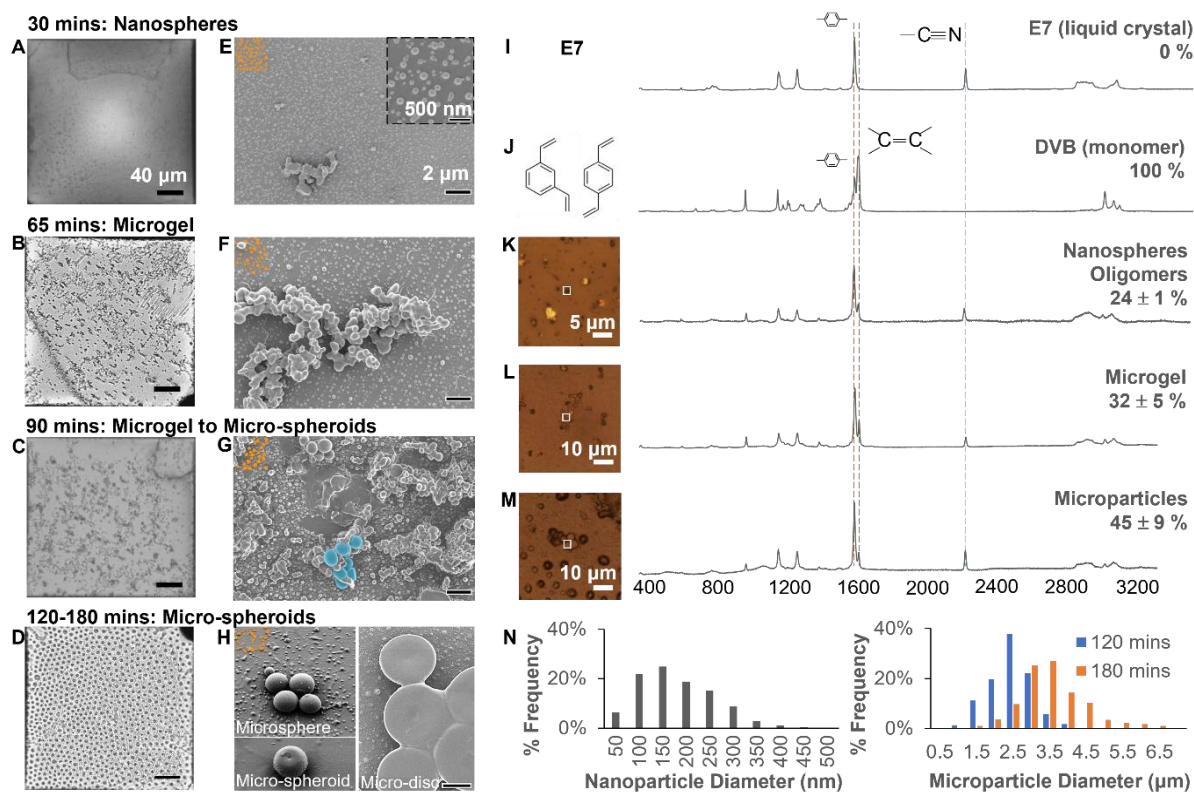

**Figure 3. Progression of morphology of pDVB particles from nanospheres to microgel clusters to micron-sized spheres, spheroids, and discs in LC on glass coated by polyimide (PI). (A-D)** Ex-situ bright field optical micrographs showing the formation of shaped polymer structures (dark clusters) in LC. **(E-H)** Representative SEM images revealing the detailed shapes of the polymer particles formed at the LC-substrate interface. A subpopulation of the precipitated nanospheres is highlighted in orange (false colored) in each SEM image to illustrate their continuous precipitation. A small number of microparticles emerging from a microgel cluster are highlighted in blue (false colored) in G to illustrate microgel cluster phase separating into discrete particles. **(I-M) Laser confocal Raman microscopy of E7, DVB monomer, and the various polymer morphologies.** Microscopy images on the left in each panel show the polymer structure on which the spectrum to the right was taken; the white box indicates the excitation area. The percent represents vinyl bonds left unreacted in each case (normalized by vinyl bonds in DVB monomer) as derived from the Raman spectra. **(N) Particle size distribution** of the nanospheres and microparticles formed by the iCVD-in-LC process.

To demonstrate that the polymer morphology evolves over time, we stopped the iCVD polymerization in LC (on glass coated by polyimide (PI)) at predetermined times (i.e., 30, 65, 90, 120, and 180 minutes). The progression of polymeric structures observed during in-situ monitoring was confirmed via ex-situ optical micrographs (**Figure 3**A-D and Supporting Information, Figure S2 and Movie S4). To characterize the microscopic progression of polymer morphology, we removed bulk LC by washing samples in ethanol and examined the LC-substrate interface at each time point using SEM. A Critical Point Drying (CPD) procedure was used to demonstrate that the ethanol washing step did not change the polymer morphology (see Supporting Information, Figure S3). The SEM images illustrate the detailed morphology progression for DVB polymerization in LC over the course of 180 minutes (Figure 3E-H). The spatiotemporal evolution of the chemical composition of the polymer structures was determined using laser confocal Raman microscopy (Figure 3I-M). As benchmarks, we first obtained the Raman spectra for nematic E7 (Figure 3I, see Supporting Information, Figure S4 for molecular structure of constituents for the E7 LC mixture) and DVB monomer (Figure 3J), which agreed well with literature.[43–47] The peaks at 1609 cm$^{-1}$ (indicated by an orange dashed line) and 2230 cm$^{-1}$ (indicated by a blue dashed line) in the E7 spectrum correspond to skeletal stretch of phenyl rings and CN stretch, respectively. We used the CN stretch (2230 cm$^{-1}$) to quantify E7 because it does not overlap with DVB peaks.[43–45] DVB monomer also has a peak at 1609 cm$^{-1}$ due to its benzene ring; the peak at 1632 cm$^{-1}$ (indicated by a green dashed line) corresponds to C=C stretch of unreacted vinyl group, which does not overlap with peaks of E7.[46,47] The C=C peak was hence used to estimate the degree of reaction of the two vinyl bonds in DVB, which was correlated to the polymeric morphology formed in the iCVD in LC process (Figure 3K-M, see Supporting Information for a detailed description of the characteristic peaks for E7 and DVB and the methodology for calculating % unreacted vinyl bonds). All shaped particles contained both E7 and pDVB.

The reaction mixture is initially homogenous, as evidenced by our in-situ observations revealing the absence of liquid-liquid phase boundary development during the continuous delivery of DVB into E7 (Supporting Information, Movie S4). Polymerization is initiated via the delivery of TBO radicals. Nanospheres are the dominant shape during the early-stage iCVD in LC polymerization (i.e., first 30 mins, Figure 3E). We attributed the formation of nanospheres to the phase separation of pDVB from the marginal solvent (i.e., E7), building upon the mechanism of precipitation polymerization of DVB in isotropic solvents.[25] The nanospheres have 24% unreacted vinyl groups (Figure 3K), which is common for precipitation polymerization.[26] However, the percentage is lower than that for pDVB synthesized via conventional iCVD (i.e., 52%),[48] which is likely a result of the low threshold concentration of DVB in nematic LC and the diffusion-limited reaction conditions. The nanoparticles have an average diameter of $160 \pm 80$ nm (Figure 3N), calculated using a sample size of 15417 particles (see Supporting Information for methodology). That size distribution was further corroborated by DLS (Supporting Information, Figure S5). This average size is an order-of-magnitude smaller than that obtained in common precipitation polymerization, i.e., 2-5 μm,[25,27,49,50] which we attribute to the combined effects of (i) the marginal solvency of pDVB in nematic E7, and (ii) the poor dispersion of the nanospheres that are larger than the extrapolation length of E7 (~100 nm),[38] owing to the elastic energy that arises from director strain.[51] (The extrapolation length of LC is defined as the ratio of a characteristic Frank elastic constant to the LC-surface anchoring energy density, which has also been postulated to determine the diameters of nanofibers synthesized in LC.[38]) This conclusion is supported by the prior observation that nanospheres with similar sizes ($300 \pm 100$ nm) have been obtained via polymerization of crosslinkers in LC.[33,52]

In contrast to precipitation polymerization, during which pDVB seed particles form mainly during early-stage polymerization, nanospheres continue to form and precipitate during the entire iCVD process in the LC, as seen from their presence at each time point (e.g., on top and around the polymeric structures that later formed, with a subpopulation false colored in orange in the upper left corners of Figure 3E-H). This observation highlights a key difference between the batch-wise reaction in precipitation polymerization, where solutions were preloaded with a finite amount of initiator and monomer, and the continuous reaction we performed with constant delivery of DVB and TBPO into nematic E7 throughout the reaction time.

We observed the precipitated nanospheres to aggregate at the LC-substrate interface, forming colloidally stable microgels after 65 minutes of iCVD polymerization (Figure 3F). Drawing an analogy to precipitation polymerization again, the precipitated nanospheres can be viewed as building blocks, i.e., monomers, which aggregate to form a microgel, analogous to an oligomer. Microgels continue to grow by further capturing nanospheres and by heterogeneous polymerization (Figure 3E-F). While precipitation polymerization often leads to spherical microgels,[25] the shape of microgels observed in Figure 3F is guided by the local orientation of the LC (see next subsection for details). These microgel clusters have 32% unreacted vinyl groups (Figure 3L), which is higher than the precipitated nanospheres and likely a result of the heterogeneous polymerization.

Similar to precipitation polymerization, discrete microparticles ultimately emerge from microgels, as we observed after 90 minutes of iCVD polymerization (Figure 3G). Nevertheless, distinct from precipitation polymerization, which typically yields uniform spherical particles, iCVD polymerization in LC achieves an unprecedented morphological diversity, including microspheres, micro-spheroids, and micro-discs at the LC-substrate interface (Figure 3H). To the best of our knowledge, spheroids and discs have not been formed previously by polymerizing DVB.

Raman microscopy revealed that pDVB microparticles have 45% unreacted vinyl groups (Figure 3M), higher than those of nanospheres and microgel clusters. This significant increase is likely due to auto-acceleration that is commonly observed in free-radical polymerization with phase separation,[53–57] including that in LC.[32,34,58] We postulate that phase separation creates a polymer-rich phase (which ultimately leads to discrete microparticles) and a solvent-rich phase, with DVB monomer preferentially partitioning into the polymer-rich phase and thus creating an environment that resembles bulk polymerization. With the high monomer concentration and high viscosity in the polymer-rich phase, chain termination and crosslinking are reduced drastically, leading to a high % unreacted vinyl groups like we observed in Figure 3M. Upon further polymerization (from 120 to 180 minutes), the microparticles grow from $2.22 \pm 0.57$ μm to $3.31 \pm 0.90$ μm in size (Figure 3N).

Our observations via the in-situ microscope, SEM, and Raman microscopy point to a strategy to achieve shape-controlled polymer particles simply by controlling the polymerization reaction time in LC. The pDVB particles undergo a pathway of nanosphere (in bulk LC) to microgel, to microsphere and micro-discs at the LC-substrate interface as the iCVD polymerization in LC progresses. We also observed the formation of a porous film at the LC-vapor interface for longer reaction times (e.g., longer than 120 mins), as shown in Supporting Information, Figure S6. Similar observations have been made during iCVD in isotropic liquids and is likely due to monomer adsorption at the LC-vapor interface.[59] Wei et al. in a recent study characterized the film morphology on different liquids (including isotropic liquids and nematic LC) in detail.[60] We refer readers to these studies to learn more about the formation of polymer thin films at the LC-vapor interface. Next, we focus on unraveling the effect of the LC anisotropy on the morphology of the polymer particles formed at the LC-substrate interface by performing iCVD in LC with three different orientations.

**Role of LC in shaping and directing DVB polymerization at LC-substrate interface**

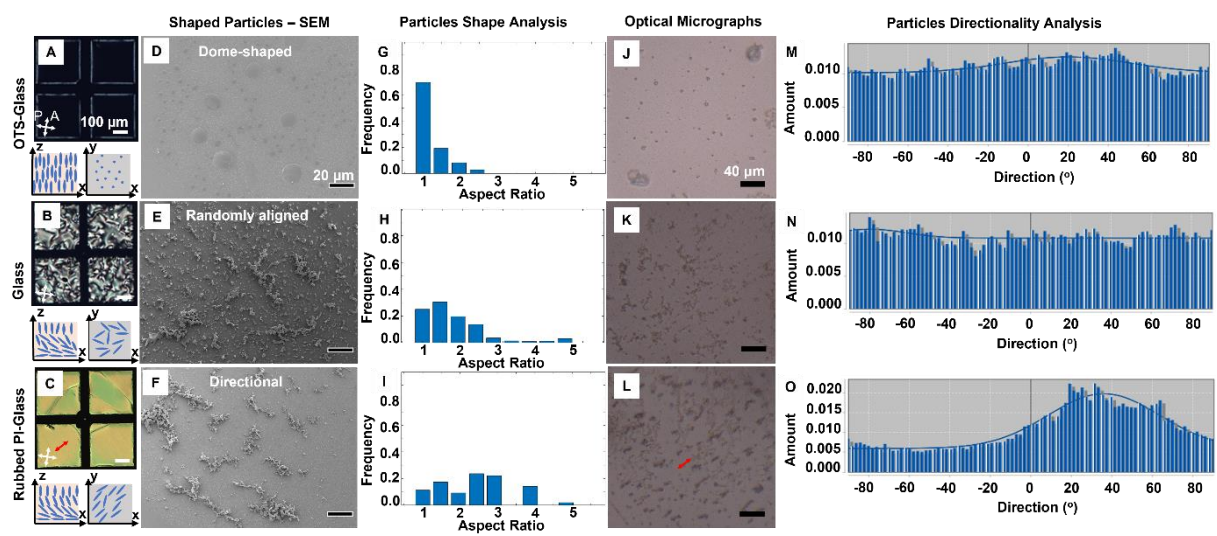

**Figure 4. Role of LC alignment in shaping and directing microgel clusters.** (**A-C**) LC alignment on the three solid substrates used in this work: glass functionalized by octadecyltrichlorosilane (OTS), bare glass, and glass coated by PI and then rubbed. In each panel, the representative cross-polarized micrographs are accompanied by the corresponding schematics showing the orientation of E7 in the x-z plane (i.e., cross-section) and x-y plane (i.e., top-down at the LC-substrate interface), respectively. The red arrow corresponds to the rubbing direction on the PI substrate. (**D-F**) SEM images of the polymeric structures formed in each LC template: OTS-treated glass, bare glass, and glass covered by rubbed PI led to dome-shaped, randomly oriented, and directional microgel clusters, respectively. (**G-I**) The frequency of appearance of microgel clusters with specific aspect ratios, revealing the statistics of shape anisotropy on the three substrates. OTS-treated glass led to circular or near-circular shape due to the homeotropic alignment of E7; bare glass and glass covered by rubbed PI led to elongated microgels due to their hybrid alignment of E7, as shown in the x-z plane orientation in A-C. (**J-L**) Bright field optical micrographs obtained for the three substrates, where polymeric clusters were captured as dark clusters. The red arrow corresponds to the rubbing direction of the PI substrate. (**M-O**) Directionality analysis of the polymeric clusters in J-L, highlighting

the aligned polymeric structures on rubbed PI (O), oriented along the presumed direction of rubbing, in contrast to the absence of directionality on OTS-treated glass and bare glass (M, N).

To understand the effect of LC anisotropy on shaping the microgels, we investigated iCVD in E7 films supported by three types of solid substrates: glass functionalized by OTS, glass, and glass functionalized with PI rubbed in one direction. E7 aligns perpendicularly at the air-LC interface. OTS functionalized glass is known to lead to perpendicular anchoring [as shown in bottom-left (x-z plane) and bottom-right (x-y plane) panels in **Figure 4**A], which was confirmed by the dark cross-polarized optical micrograph (Figure 4A top). The glass surface (cleaned with alconox and ethanol prior to use) leads to planar anchoring with random azimuthal orientation, and a hybrid orientation in the cross-section (Figure 4B). Rubbed PI-functionalized glass leads to planar anchoring with uniform azimuthal orientation of E7 along the direction of rubbing, which was also confirmed using cross-polarized optical imaging (Figure 4C).[42]

The iCVD polymerization undergoes similar growth stages in LC that is supported by OTS-functionalized glass, glass, and PI-functionalized glass (with the latter described in previous subsection; see Supporting Information Figure S7-10 for a detailed discussion on polymer morphology progression during iCVD in LC on the other two surfaces). However, intriguingly at the interfacial microgel stage, the LC anisotropy at the LC-substrate interface shaped the microgel clusters, while translating its long-range molecular order into the ordering of the clusters.

The SEM images of the microgels formed on the three substrates (Figure 4D-F) clearly indicate the directed growth of disparate polymer morphology. That apparent shape anisotropy is quantitatively demonstrated through a statistical analysis of the frequency of appearance of the microgel clusters with distinct aspect ratios on each substrate (detailed methodology is described in Materials and Methods section). Dome-shaped hemispheres (aspect ratio = 1) emerge on OTS-functionalized glass (Figure 4D, 4G) due to the homeotropic E7 alignment. In contrast, elongated particles with higher aspect ratios dominate in E7 with hybrid alignment, i.e., on glass and rubbed PI-functionalized glass. Specifically, unmodified glass leads to randomly aligned microgels (Figure 4E, 4H), whereas directional microgel (oriented along the presumed direction of rubbing) forms on a rubbed PI surface (Figure 4F, 4I).

Based on this observation at the micron-scale, we performed large-scale particle directionality analysis at the length scale of a few millimeters, on polymer clusters observed in optical micrographs (Figure 4J-L, detailed methodology is described in Supporting Information). The directionality histograms show that the microgels formed on OTS-functionalized glass and glass are randomly aligned (Figure 4M, 4N), and those formed on rubbed PI are aligned at an angle of 36º (Figure 4O), i.e., the direction of rubbing. This directionality is further seen through the continuous alignment of microgels over the length scale of 3.05 mm (see Supporting Information, Figure S11).

**Synthesis capabilities and potential applications**

Building upon the mechanistic understanding of LC-templated iCVD polymerization, we achieved a host of polymer morphologies simply by tuning the reaction time and LC alignment (Figure 5A). By limiting the polymerization time to 30 minutes while using a homeotropic or a hybrid LC alignment, we obtained predominantly nanoparticles. At 65 minutes, dome-shaped particles were obtained in homeotropic LC, whereas elongated microgels were obtained in LC with hybrid alignments. Microparticles with different shapes (spheres, spheroids and, discs) were synthesized by performing iCVD for 120-180 minutes in LC with hybrid alignment. Each polymer morphology is associated with its characteristic feature size (e.g., the diameter of nanoparticles is ~160 nm and that of microparticles is 2-3 μm).

The iCVD conditions (i.e., monomer and initiator flow rates, total chamber pressure, stage and temperatures) were kept constant in these depositions (see Supporting Information, Table S1 for detailed deposition conditions). Engineering these deposition conditions will, we predict, lead to a greater morphological diversity, which will be a focus of our future studies. Furthermore, iCVD is compatible with a library of more than 70 monomer chemistries, which can be harnessed to uncover the endless combinations of morphology and organic functionality that may be accessible using the reported approach.

These shape-controlled particles have many advantages in a broad range of applications. Here we briefly demonstrate their potential as drug delivery vehicles. The E7 encapsulated in the particles (as indicated by the CN peak at 2230 cm$^{-1}$ in their Raman spectra, Figure 3K-M), can be viewed as a proxy for therapeutics. To demonstrate the ability to release similar molecules, we studied the Raman spectra for micron-sized spheroid particles before and after immersing in ethanol for 10 mins. Disappearance of the CN peak at 2230 cm$^{-1}$ after release (Figure 5B) indicates successful release of the encapsulated E7 molecules with no change in the polymer composition (i.e., % unreacted vinyl bonds) (see Supporting Information, Figure S12). Although the release conditions will likely need to be modified to adapt the particles to biomedical applications, we believe this proof-of-principle points to their exciting future potential. We also note the heterogeneity in particle sizes shown in Figure 5A and 5B, which can be desirable for programming the pharmacokinetics using a carefully selected collection of such particles.

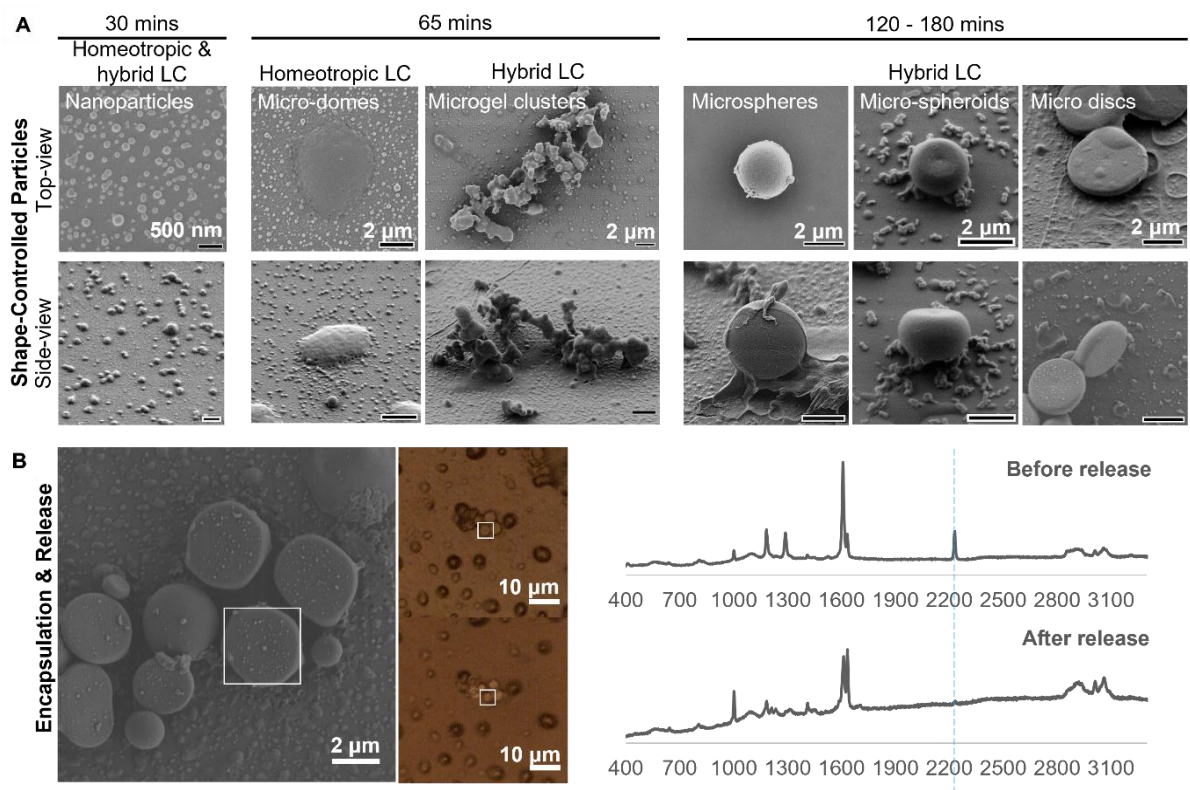

**Figure 5. (A) Synthesis of shaped polymeric particles using iCVD in LC.** SEM images of the particle shapes (top & side views) that show the detailed polymer morphology at different growth stages. Although different samples were examined in the top- versus side-view SEM images, they were synthesized under identical conditions, reaction time, and LC template. Nanoparticles are synthesized by performing reaction for 30 minutes in both homeotropic and hybrid LCs; hemispherical domes and elongated clusters are formed by performing reaction for 65 minutes in homeotropic and hybrid LC respectively; microparticles with different shapes (spheres, spheroids, discs) are formed by performing reaction between 120-180 minutes in hybrid LC. **(B) Encapsulation and release of E7 molecules by the shape-controlled particles.** SEM image (left) of the corresponding particle for which the Raman spectra (right) were collected and, microscopy image (middle) and the Raman spectra (right) obtained before and after release of E7, respectively.

## DISCUSSION

We uncovered a two-step pathway governing the synthesis of polymeric structures in an anisotropic medium, which led to disparate morphologies, using DVB as the monomer for demonstration. Drawing inspiration from the well-understood mechanisms of DVB precipitation polymerization in isotropic solvents, we revealed that LC-templated iCVD polymerization goes around the precipitation growth pathway twice, once in bulk and the second time at the LC-substrate interface. This two-step pathway provides an effective framework for understanding the mechanism of polymerization in anisotropic solvents. It also enabled us to achieve a host of polymer morphologies previously inaccessible in isotropic solvents. The templating effects of LC were demonstrated using three types of LC alignments, carefully selected to have different orientation near the LC-substrate interface, which enabled us to imprint LC anisotropy on the polymeric microgels, an initial step towards shape programmability.

Those insights were enabled in part by the technical advancement of in situ and real-time monitoring of the iCVD polymerization in LC. An in-situ reflectance microscope serves as an in-line monitoring technique to obtain the targeted polymer shapes while maintaining the anisotropy of the medium. The obtainment of targeted shapes also requires meticulous tuning of the reaction conditions, which was enabled by the precision synthesis capability of iCVD, such as continuous vapor reactant delivery with precise compositions of initiator and monomer.

With the expansive library of polymer chemistries that has been demonstrated using the iCVD technique, the reported approach of iCVD in LC and its fundamental mechanisms will open many future opportunities. Considerable future work is required to understand the effects of monomer and polymer solubilities on shape formation, of the reaction conditions (such as initiator concentration and type, substrate surface energy, other anisotropic templates, to name a few) on the polymerization mechanism, rate of diffusion, and the resultant polymer morphology. For example, the emergence of similar micro-spheroids and micro-discs has been observed during the phase separation of a polystyrene particle and decane molecules absorbed into the particle (the phase separation was driven by their solubility change during cooling).[61] Alternatively, those particle shapes can form through buckling of microspheres driven by a capillary pressure across the phase boundary due to the diffusion of LC out of the polymer-rich phase.[62–64] An important focus of our future investigation will be to fully elucidate the mechanism of shape formation.

This work demonstrates that iCVD is a versatile tool for the synthesis, characterization, and mechanistic investigation of polymerization templated by structured liquids. Its compatibility with established manufacturing technologies, such as roll-to-roll semicontinuous deposition, makes it an attractive approach for the design and manufacturing of functional polymeric particles with targeted shapes. That novel capability will benefit many existing and future technologies, ranging from shape-programmed pharmacokinetics in drug delivery to shape-controlled motion and group behavior in fluids for soft robotics.

## MATERIALS AND METHODS

### Materials

We obtained the nematic LC, E7 (nematic mixture of cyanobiphenyls and terphenyls), from EMD Millipore (Billerica, MA). Divinylbenzene (technical grade, 80%), hexane (CHROMA SOLV ® grade) and octyltrichlorosilane (OTS), were purchased from Sigma-Aldrich (St. Louis, MO) and used as received. Polyimide and Thinner solutions were purchased from HD MicroSystems (Parlin, NJ). Ethanol (anhydrous, 200 proof) was purchased from VWR International (Radnor, PA) and used as received. TEM copper grids and quartz slides were purchased from Electron Microscopy Sciences (Hatfield, PA).

### LC templates preparation

Substrate Preparation: Glass substrates were prepared by incubating a glass slide in alconox solution and sonicating for 10 minutes, followed by a thorough rinse in deionized water (DI-H2O) and ethanol. To functionalize glass substrates with OTS, they were first incubated in a hexane solution of OTS (324uL in 140mL hexane) for 30 min followed by rinsing with chloroform for 1 min and were then thoroughly rinsed in deionized water (DI-H2O) and ethanol. To functionalize glass substrates with PI, polyimide (PE2555) and thinner solutions were mixed (1:1 wt/wt) in a vial and vortexed for 30 seconds. The mixture was filtered with a 0.45-μm pore size filter. Glass substrates (rinsed with ethanol) were then spin coated with the filtered PI + thinner mixture (3000RPM for 25 seconds). This was followed by baking the surfaces at 275C for 1 hr and then rubbing them unidirectionally with a velvet cloth.

Preparation of LC Templates: A TEM grid (18 μm in thickness) was placed on the above-described substrates and filled with E7. Excess LC was then removed from the grid using a capillary tube, yielding an 18 μm thick LC layer for all the above mentioned systems.

**Characterization Methodologies**

Microscopy: An Olympus BX41 microscope with 4X and 20X objectives, two rotating polarizers, and a Moticam 10.0 MP camera was used for optical microscopy. In addition to crossed polarizers in the above microscope setup, a Bertrand lens was used for conoscopic polarized light microscopy. For in-situ microscopy, a custom setup with Keyence VHX 970F long-focal length (85mm) microscope lens (50x to 500x magnification) was installed directly over iCVD reactor chamber. Scanning electron microscopy (SEM) images were captured at an acceleration voltage of 2 kV on samples that had been coated with approximately 3 nm of gold/palladium using a Zeiss LEO 1550 field emission scanning electron microscope (FESEM).

Dynamic Light Scattering (DLS): One mL ethanol solutions of the polymers prepared by iCVD-in-LC were placed in glass tubes and sonicated for 30 minutes. Dynamic light scattering measurements (Brookhaven Instrument) at room temperature were recorded at light scattering angles 90° using a laser of wavelength 638 nm and the corresponding autocorrelation functions were generated. The autocorrelation functions were analyzed using the CONTIN algorithm to calculate the hydrodynamic diameters of the polymer particles in the ethanol solution.

Critical Point Drying (CPD): Since solvent evaporation induce capillary forces that can affect the shape of the particles, we employed critical point drying procedure to scrutinize the effect of post processing washing step using external solvents (eg. ethanol, isopropanol) used by us to remove the LC. In this method, the iCVD-in-LC polymerized samples were immersed in a bath of isopropanol which was then replaced by liquid carbon dioxide using a LEICA CPD300. The isopropanol substitution by liquid $CO_2$ was performed for 12 cycles. The $CO_2$ was then removed at the critical temperature (31°C and 74 bar) to avoid the generation of a meniscus and capillary forces acting on the polymer network owing to similar density of liquid and its vapor phase.

Raman microscopy: Raman spectra were collected on samples prior to SEM using Renishaw InVia Confocal Raman imaging microscope. All spectra were recorded using excitation with a 532 nm laser at 40mW with a spatial resolution ~ 1μm and spectral resolution of ~ 1 $cm^{-1}$ using 2400 lines/mm grating and an edge filter with CCD camera detector. Two objectives were used, 50x (N.A. 0.45 for Raman imaging, long-working distance with estimated axial resolution ~ 2.6 μm) and 100x (N.A. 0.9 for Raman imaging with estimated axial resolution ~ 0.7 μm).

Spectral acquisition conditions used were as follows:

| Sample type | Raman shift ($cm^{-1}$) | Exposure time (sec) | Laser power (%) | Shutter (%) | Number of Scans |
|---|---|---|---|---|---|
| DVB | 400 – 3600 | 10 | 10 | 10 | 3 |
| E7 | 400 – 3600 | 10 | 50 | 50 | 3 |

| Qualitative analysis spectra for particles | 300 - 3300 | 10 | 50 | 50 | 3 |
| Quantitative analysis spectra for particles | 1340 - 2400 | 10 | 50 | 50 | 10 |

For qualitative composition analysis, conditions were optimized to capture the entire spectral range to identify all peaks for individual particles, E7 and DVB without detector saturation. To capture individual nanoparticles, 100x objective was used while 50x (long-working distance) was used to capture individual micron-sized particles. However, to capture all the data used for quantitative comparisons, 50x (long-working distance) objective was used, with same spectral acquisition parameters as mentioned above. This was to ensure the same spatial resolution (both lateral & axial) to capture the evolution of chemical compositions of particles synthesized throughout the shape progression in iCVD-in-LC polymerization at LC-substrate interface. This implied that the properties of individual micron-sized particles were captured, however, for the nanospheres composition of a group of nanoparticles combined with heterogenous polymeric chains formed at LC-substrate interface were captured.

**Raman Spectra Peak Analysis and Calculation of % Unreacted Vinyl Bonds**

Signature Peaks: As benchmarks, we first obtained the Raman spectra for nematic E7 and DVB monomer, which agreed well with literature.[43–47] Briefly, peaks at 1184 and 1609 cm$^{-1}$ correspond to aromatic CH in-plane deformation and C=C skeletal stretch of phenyl rings of E7, respectively. The peaks in the region of 2850-2950 cm$^{-1}$ are from the aliphatic chains attached to the benzene ring in E7 and peaks 3010-3100 cm$^{-1}$ are due to CH stretch of benzene ring. The peak at 1288 cm$^{-1}$ is due to C-C stretch of biphenyl bond and peak at 2230 cm$^{-1}$ due to CN stretch, which is the signature peak used here since it is proportional to number of E7 molecules and it occupies a region without any overlap with the DVB peaks.[43–45] Prominent peaks for DVB monomer correspond to breathing vibrational mode and skeletal stretch of benzene ring at 1000 and 1609 cm$^{-1}$, respectively. The peaks in the range 1150-1200 cm$^{-1}$ correspond to C-C stretches and in the range of 3000-3100 cm$^{-1}$ belong to C-H stretches. The peak at 1632 cm$^{-1}$ correspond to the C=C stretch of unreacted vinyl group which is of importance to us to calculate the percentage of the unreacted vinyl bonds in pDVB.[46,47] This is also the signature peak of DVB that does not overlap with peaks of E7. In summary, the three peaks used to estimate unreacted vinyl bonds of polymer particles are: the nitrile peak (2230 cm$^{-1}$) in E7, C=C stretching of unreacted vinyl bonds in DVB (1632 cm$^{-1}$), and ring skeletal stretch that are shared by E7 and DVB (1609 cm$^{-1}$).

Quantitative Analysis: To quantify the % of unreacted vinyl groups, we deconvoluted the peaks at 1583, 1609 and 1632 cm$^{-1}$ and calculated the peak areas for the three peaks of importance at 1609, 1632 and 2230 cm$^{-1}$ using Origin data analysis software. For that calculation, we first plotted the spectra from obtained data, followed by software guided peak analysis which included background subtraction (using an interpolated user defined baseline by identifying 8-10 anchor points on the baseline of spectra), peak deconvolution and area integration. Next, using the ratio of peak areas corresponding to phenyl ring (at 1609 cm$^{-1}$) and CN stretch (at 2230 cm$^{-1}$) obtained for E7 as reference, we deconvolute the contribution to peak area (of 1609 cm$^{-1}$) of phenyl rings in E7 and pDVB for all polymeric structures. Taking the ring to unreacted vinyl peak area ratio in DVB spectra as reference, we then calculate the percentage of vinyl groups left unreacted in each particle using the ratio of the peak areas corresponding to rings of pDVB (deconvoluted in previous step from peak at 1609 cm$^{-1}$) to unreacted vinyl (at 1632 cm$^{-1}$).

**Image Processing and Statistical Analysis**

False coloring in Figure 3E-H was done using Adobe Illustrator by manually defining particle boundaries and overlaying a colored mask on top of the nanoparticles (orange) and microparticles (blue) present in a section of SEM images. Nanoparticles size distribution in Figure 3N was performed using FIJI. The circular cross section of 15417 nanoparticles were identified by creating binary masks from SEM images using thresholding image segmentation. To remove salt and pepper noise, despeckle function was used (which is a median filter), followed by fill holes and watershed functions (to facilitate separate detection of aggregated particles) built-in FIJI. The diameters were calculated from the particle areas calculated using Analyze Particles in FIJI. Micro-particles size distribution in Figure 3N and S11A-B was calculated by measuring the particle diameters manually for each particle in FIJI for a total of 1697 particles.

To characterize microgel polymeric structures formed using iCVD-in-LC process in different LC templates (Figure 4D-I), we quantified particle shapes by calculating the frequency of appearance of microgel clusters with specific aspect ratio by analyzing SEM images capturing microgels (1087 clusters) formed over larger grid area using FIJI. Here, Bandpass Filter was used to filter large structures down to 100 pixels and small structures up to 5 pixels to correct for the shadow effect and noise. We then created the binary masks, followed by fill holes and watershed functions (to facilitate separate detection of aggregated particles) built-in FIJI. The aspect ratio of particles was then calculated from the fitted ellipse mask using Analyze Particles plug-in in FIJI, as shown in figure S13 (SEM image, binary mask, mask after processing and fitted ellipse mask). We then created the weighted frequency plots (weighted by area) using an existing MATLAB function 'Generate Weighted Histogram' created by Mehmet Suzen (2021), (https://www.mathworks.com/matlabcentral/fileexchange/42493-generate-weighted-histogram), MATLAB Central File Exchange.

We further characterized the directionality of the microgels formed in different LC templates (Figure 4J-O) by analyzing the bright-field optical micrographs obtained over entire grids using FIJI. Here, the binary masks were created using color thresholding after correcting the captured optical micrographs for optimal brightness/contrast and noise (by using a median filter of radius 3), as shown in figure S14. Directionality histograms were created using the Fourier components method (Nbins = 90) in the directionality plug-in.

Two sample t Test (with Welch Correction) was performed using Origin for determining the p values in Figure S9, S10 and S12.

**Acknowledgement**s

This material is based upon work supported by the National Science Foundation (NSF) through the Future Manufacturing Research Grant under Grant No. CMMI-2229092 and through the Faculty Early Career Development Program under Grant No. CMMI-2144171. Analytical methods involved use of the Cornell Center for Materials Research (CCMR) Shared Facilities which are supported through the NSF MRSEC program (DMR-1719875). The authors also thank Dani Streever for her help with generating micro spheroid size distribution from collected data and Philip Carubia for his help with optimizing Raman spectral acquisition conditions.